\documentclass[preprint,5p]{elsarticle}
\usepackage{lineno,hyperref}
\usepackage{array}
\modulolinenumbers[5]
\usepackage{siunitx}

\journal{Nuclear Instruments and Methods in Physics Research Section A}

\makeatletter
\def\ps@pprintTitle{%
 \let\@oddhead\@empty
 \let\@evenhead\@empty
 \def\@oddfoot{\footnotesize\itshape Prepared for submission to \ifx\@journal\@empty Elsevier \else\@journal\fi\hfill\today}%
 \let\@evenfoot\@oddfoot}
\makeatother

\begin{document}

\begin{frontmatter}

\title{Ultra-low background mass spectrometry for rare-event searches}

\author{J.~Dobson\corref{cor1}}
\ead{j.dobson@ucl.ac.uk}
\author{C.~Ghag, and L. Manenti\corref{}}
\cortext[cor1]{Corresponding author}

\address{Department of Physics and Astronomy, University College London, Gower Street, London, WC1E 6BT, UK}

\begin{abstract}
Inductively Coupled Plasma Mass Spectrometry (ICP-MS) allows for rapid, high-sensitivity determination of trace impurities, notably the primordial radioisotopes $^{238}$U and $^{232}$Th, in candidate materials for low-background rare-event search experiments. We describe the setup and characterisation of a dedicated low-background screening facility at University College London where we operate an Agilent 7900 ICP-MS. The impact of reagent and carrier gas purity is evaluated and we show that  twice-distilled ROMIL-SpA\textsuperscript{TM}-grade nitric acid and zero-grade Ar gas delivers similar sensitivity to ROMIL-UpA\textsuperscript{TM}-grade acid and research grade gas. A straightforward procedure for sample digestion and analysis of materials with U/Th concentrations down to 10 ppt g/g is presented. This includes the use of $^{233}$U and $^{230}$Th spikes to correct for signal loss from a range of sources and verification of $^{238}$U and $^{232}$Th recovery through digestion and analysis of a certified reference material with a complex sample matrix. Finally, we demonstrate assays and present results from two sample preparation and assay methods: a high-sensitivity measurement of ultra-pure Ti using open digestion techniques, and a closed vessel microwave digestion of a nickel-chromium-alloy using a multi-acid mixture. 
\end{abstract}


\end{frontmatter}


\section{Introduction}
\label{sec:introduction}
Rare event search experiments such as those seeking interactions from galactic dark matter scattering or evidence of neutrino-less double beta decay ($0\nu\beta\beta$) require construction from intrinsically radio-pure materials to limit radiogenic backgrounds. Experiments must conduct radio-assay campaigns to select materials and to accurately characterize residual radioactivity for the experiment's background model against which any observed excess will be evaluated and statistical confidence ascribed. The radioactive isotopes $^{238}$U and $^{232}$Th and their decay-chain progeny are of particular concern, contributing the bulk of the $\gamma$-ray, $\alpha$-particle, and, via spontaneous fission and ($\alpha$,n) reactions, neutron-induced backgrounds.  

Comprehensive radio-assay campaigns now increasingly deploy multiple techniques to build a high-precision background model, and to meet sample throughput requirements during the experiment's construction phase. Mass spectrometry techniques such as ICP-MS can be used to directly measure trace quantities of $^{238}$U and $^{232}$Th, delivering assays from small samples that are digested and screened with turnaround times on the order of 1--2 days. ICP-MS can achieve part-per-trillion (ppt) g/g level sensitivity to $^{238}$U and $^{232}$Th when sample preparation protocols with stringent cleanliness constraints are used to limit contamination~\cite{ref:thinkblankbook, doi:10.1021/acs.analchem.6b04854, Leonard:2014jha}. Gamma-spectroscopy is used to assay larger samples, including finished components, over several weeks to determine the activities of the decay chain daughters~\cite{Scovell:2017srl}. Other techniques such as Glow-Discharge Mass Spectrometry (GD-MS) and Neutron Activation Analysis (NAA) may also be used for some materials not well-suited to either ICP-MS or gamma-spectroscopy.  

Here we present results from a new ICP-MS facility at University College London (UCL) constructed to perform radio-assays in support of rare-event searches, such as the `Generation-2' LUX-ZEPLIN (LZ) dark matter Experiment~\cite{Mount:2017qzi}. In Section~\ref{sec:lab_overview} we give an overview of the UCL ICP-MS facility and investigate the impact of reagent and Ar gas purity on achievable sensitivity. Then, in Section~\ref{sec:procedure_and_analysis}, we describe a straightforward procedure for routine and fast-turnaround of materials with U/Th concentrations down to 10 ppt g/g. We present detection limits for different vessel cleaning protocols and describe the use of $^{230}$Th and $^{233}$U spikes for signal loss correction. A certified reference material representing a realistic sample matrix is processed to verify overall U/Th signal recovery. Finally, in Section~\ref{sec:ti_and_inconel_assays} we present results from an open digestion of ultra-pure Ti and a closed vessel digestion of a nickel-chromium-alloy requiring multi-acid digestion. 

\section{Laboratory Overview}
\label{sec:lab_overview}

The UCL ICP-MS facility was installed and commissioned in a newly constructed ISO Class 6 cleanroom in late 2015. The primary instrument is an Agilent 7900 ICP-MS with an instrument sensitivity to U and Th down to $10^{-15}$--$10^{-14}$ g/g (1--10 parts-per-quadrillion (ppq) g/g). The ICP-MS is fitted with an integrated auto-sampler for high-speed discrete sample uptake with a low-flow, Peltier-cooled sample introduction system. The system octopole can provide species discrimination through interference removal by running in either the He collision or H$_{2}$ reaction gas mode. High-purity Ar (N5.5/N5.0), He and H$_{2}$ (N6.0) are introduced to the system as carrier, collision and reaction gasses, respectively. The system is fitted with an inert sample introduction kit that includes a PFA nebulizer, spray chamber and torch connector, platinum sampling and skimmer cones and a plasma torch with quartz outer body and sapphire injector. This allows up to $20$\% v/v acid concentration in the sample introduced to the ICP-MS, including hydrofluoric acid (HF). 

The limiting factor in realising reproducible high throughput ppt sensitivity is clean, standardised sample preparation requiring dedicated digestion apparatus and well developed procedures, including the use of ultra-pure acids to avoid contamination of samples. The facility has the relevant infrastructure to provide this, including: sample preparation in separate ISO Class 5 laminar flow unit (LFU), sample digestion with the Milestone ETHOS--UP closed vessel microwave digestion system with SK--15 high pressure rotor, Milestone sub-boiling point acid distillation (subCLEAN) and reflux cleaning (traceCLEAN) systems and an ELGA PURELAB flex ultra-pure de-ionised (DI) water supply (18.2 M$\Omega$.cm, $<5$ parts-per-billion (ppb) total organic carbon). Finally, a Pyro--260 microwave ashing system allows for the digestion of materials such as PTFE that are resistant to most acids, including HF.

For most assays we follow material-by-material digestion routines available for the ETHOS--UP and SK--15 high pressure rotor. These specify the acid combinations and microwave heating profiles needed to achieve complete digestion. Where these do not exist, custom routines for non-standard materials have been developed based on existing routines and in partnership with Milestone and U.K. supplier Analytix Ltd. Typically nitric acid (HNO$_{3}$) is used to digest organic material, hydrofluoric acid (HF) for decomposing silicates and metals such as Ti that are resistant to oxidising acids, and hydrochloric acid (HCl) for Fe-based alloys. When fitted with the SK--15 high pressure rotor the ETHOS--UP can simultaneously digest up to 15 samples (with the same chemistry) each in their own 100 mL TFM vessels. TFM is a chemically modified form of PTFE with excellent properties for trace analysis: high chemical resistance, low-trace metal impurities, high melting point and an extremely smooth surface for decontamination and cleaning.  

\subsection{Reagent purity}

The purity of acids and other reagents used for digestion, vessel-cleaning and preparation of calibration standards is critical for the limit of detection achievable when assaying a material. For standard cleaning of vessels and digestion of samples with $\gg\mathrm{ppb}$ U/Th concentrations we use ROMIL--SpA\textsuperscript{TM} Super Purity Acids (SpA). These are cost effective acids produced using sub-boiling point distillation and are typically certified to $< 0.1$~ppb $^{238}$U and $^{232}$Th per gram of reagent. For lower detection limits and sub-ppb materials we use ROMIL-UpA\textsuperscript{TM} acids and reagents which are produced through multiple re-distillation and are typically certified to $< 0.1$ ppt $^{238}$U and $^{232}$Th.  

Table~\ref{tab:Reagent_BECs} compares measured background equivalent concentrations (BEC) for nitric acid process blanks made with ROMIL--SpA and ROMIL--UpA acids as well as for ROMIL--SpA nitric acid that was subsequently twice-distilled using the in-house subCLEAN distillation system. Inferred concentrations for undiluted HNO$_{3}$ are consistent with the manufacture's stated typical concentrations: $< 100$ ppt U/Th for SpA and $< 0.1$ ppt U/Th for UpA. We find that twice-distilled SpA is consistent with the UpA background concentration for $^{232}$Th and approaches that for $^{238}$U. Based on these results we now use twice-distilled SpA--grade nitric for most digestions unless ultimate detection limits are required. In addition, twice-distilled SpA can be used in cleaning procedures requiring larger quantities of reagents where UpA--grade acid is prohibitively costly. 

\begin{table}[h]
\centering
\caption{$^{232}$Th and $^{238}$U BECs for different grades of ROMIL nitric acid diluted to $\sim20$\% v/v with DI water.  Solutions were prepared in 50 mL PP (polypropylene) vessels pre-leached in SpA nitric acid for 24 hours. Equivalent concentrations for concentrated 69\% nitric acid are shown. These represent upper limits on the acid purity as contributions from other sources, such as leaching from vessels, are not subtracted.}
\medskip
\begin{tabular}{ 
    r |
    S[table-format=1.3]@{\,\( \pm \)\,}
    S[table-format=1.3]
    S[table-format=1.2]@{\,\( \pm \)\,}
    S[table-format=1.2] 
  }
  \multicolumn{1}{c}{} & \multicolumn{4}{c}{ppt g/g $\sim20$\% v/v diluted} \\
  & \multicolumn{2}{c}{$^{232}$Th} & \multicolumn{2}{c}{$^{238}$U} \\ 
  \hline
  ROMIL--SpA & 1.730 & 0.160 & 1.85 & 0.06 \\
  ROMIL--UpA & 0.035 & 0.005 & 0.03 & 0.01 \\
  distilled--SpA & 0.030 & 0.004 & 0.05 & 0.01 \\
  \multicolumn{1}{c|}{} & \multicolumn{4}{c}{Eqv. ppt g/g 69\% HNO$_{3}$} \\
  ROMIL--SpA & 6.61 & 0.60 & 7.06 & 0.25 \\
  ROMIL--UpA & 0.12 & 0.02 & 0.10 & 0.03 \\
  distilled--SpA & 0.11 & 0.02 & 0.19 & 0.04 \\
\end{tabular}
\label{tab:Reagent_BECs}
\end{table}%

\subsection{Ar gas supply}

In analysis mode the ICP-MS consumes up to 20 litres per minute of argon gas, one 50 L 200 bar cylinder for every 8 hours of running\footnote{A liquid argon supply was considered but was less cost effective due to wastage during periods of downtime.}. Given the importance of Ar gas purity for ultra-trace analysis and the potential costs associated with the high-turnaround we have compared the performance of the ICP-MS with two different grades of bottled gas: N5.0/zero-grade (min. 99.999\%) and N5.5/research-grade (min. 99.9995\%), where the cost of research-grade is significantly higher, between 5 and 10 times that of zero-grade from our supplier. 

Table~\ref{tab:Ar_checks} shows the $^{232}$Th and $^{238}$U BECs for a 5\% v/v HNO$_{3}$ acid blank for both zero- and research-grade argon. Two sets of acid blanks were prepared, a first using unconditioned labware (virgin PP DigiTubes and 1 L Nalgene containers) that was simply rinsed with DI water prior to use and a second set where all labware was first leached in 10\% v/v SpA HNO$_{3}$ for 24 hours prior to use. Switching to blanks prepared using pre-conditioned labware halved the BECs down to 7.1 ppq $^{232}$Th and 5.9 ppq $^{238}$U when using research-grade argon. Zero-grade argon shows a few ppq increase with respect to research-grade although results are almost within measurement errors. 

Switching to zero-grade did not cause deterioration in the CPS/ppt instrument response or the level of oxides  and doubly charged species during tuning; in this case (156CeO$^{+}$/140Ce$^{++}$) and (70Ce$^{++}$/Ce$^{+}$). Given this and the sub-dominant blank rate increase we conclude zero-grade argon is sufficient for most routine samples where $^{232}$Th and $^{238}$U are the analytes of interest.   

\begin{table}[h]
\centering
\caption{HNO$_{3}$ blank rates for different Ar gas supplies using 5\% v/v HNO$_{3}$ (ROMIL UpA). For the second two measurements all labware (PP vials and DigiTubes) were leached in 10\% ROMIL SpA HNO$_{3}$ for 24 hours prior to use. For each run the instrument response based on external calibrations were $\sim 300$ CPS/ppt.}
\medskip
\begin{tabular}{ 
    r |
    S[table-format=1.1]@{\,\( \pm \)\,}
    S[table-format=1.1]
    S[table-format=1.1]@{\,\( \pm \)\,}
    S[table-format=1.1]
  }
  \multicolumn{5}{r}{ppq response 5\% HNO$_{3}$ blank} \\
  & \multicolumn{2}{c}{$^{232}$Th} & \multicolumn{2}{c}{$^{238}$U} \\
 \hline
 N5.5 Ar & 13.5 & 1.4 & 11.8 & 1.8 \\  
 N5.5 Ar (leached) & 7.1 & 0.9 & 5.9 & 1.9 \\
 N5.0 Ar (leached) & 9.1 & 2.2 & 9.2 & 2.3 \\
\end{tabular}
\label{tab:Ar_checks}
\end{table}%

\section{Procedures and analysis}
\label{sec:procedure_and_analysis}

With a focus on high-throughput and fast-turnaround of time-critical assays, straightforward cleaning, sample preparation and analysis procedures have been developed for materials with U/Th concentrations in the 10~ppt to 1~ppb range: Samples are microwave-digested in preâ-cleaned TFM vessels using ultra-high purity acids. They are then diluted, without further chemical treatment, into disposable 50 mL polypropylene (PP) vessels ready for ICP-MS analysis. Fractional recoveries of $^{230}$Th and $^{233}$U spikes added prior to digestion are used to correct for $^{232}$Th and $^{238}$U signal loss from a range of sources. In particular, this enables accurate analysis of samples with high total dissolved solids (TDS) where the instrument response degrades throughout the run. A full assay including digestion, ICP-MS measurement and analysis can be completed in a single day.  

For a typical material three samples and three process blanks are  prepared\footnote{Plus a sacrificial sample in digestion vessel one, the reference vessel with a temperature probe used for microwave process control.}. Where possible, samples are weighed directly into the 100 mL TFM vessel on an analytical balance---located in the Class 5 LFU---for a differential measurement. If this is not possible the average of three separate measurements is taken before and after addition of the sample. For a 0.1 g sample this results in a systematic uncertainty of between 2-4\%. If applicable, $^{230}$Th and $^{233}$U spike solution is then pipetted into the vessel where the amount added is determined as a differential measurement with a fractional error $< 0.2$\%. The digestion vessel is then transferred to the fume cupboard where reagents are added.

Following digestion, the content of each vessel is decanted and rinsed into a 50 mL PP vessels (pre-weighed and labeled) using DI water and diluted up to a total volume of 50 mL. The PP vessels are then weighed on the analytical balance to determine overall dilution and sample concentration. If further dilution is needed to reach a sample concentration of 0.2\%\footnote{To ensure TDS are below the maximum recommended by the manufacturer to avoid significant matrix effects and contamination of sampler and skimmer cones.} or because the sample is expected to have a high ($>$ few ppb U/Th content) then a second stage of dilution into a new PP vessel is performed. The samples are then ready to be poured into 6 mL PP vials and loaded into the integrated auto-sampler.

The ICP-MS plasma auto-tune mode is chosen based on the sample type: for samples with high TDS (0.01-0.2\%) the General Purpose mode is used, whereas for low TDS the Low Matrix mode is used. The ICP-MS is then auto-tuned for high mass sensitivity using a tuning solution containing Li, Y, Tl, Co and Ce in 2\% HNO$_{3}$. The samples are then analysed in a single ICP-MS run: for routine analysis we use a batch with external calibrations, whereas for high precision measurements of samples with complex matrices we use the method of standard additions to prepare calibrations in the sample matrix itself. To monitor instrument stability throughout the batch $\sim1$~ppb of $^{193}$Ir and $^{209}$Bi internal standards are added into the sample line using a mixing tee; these are not used for corrections. Masses ($\it{m/z}$) 230, 232, 233 and 238 are monitored in single point peak mode (a single channel per mass at the mass peak) with the total integrated sample time/mass chosen based on the expected signal size: for measurements close to the detection limits greater than 20 seconds. 

All data is exported from the MasssHunter software in uncalibrated form: counts per second (CPS) plus instrument relative standard deviation (RSD) for each sample measurement and mass point. Calibration, blank rate subtraction, sample dilution and spike recovery corrections are then applied offline. The RSD errors are treated as gaussian and combined with systematic errors associated with the blank rate subtraction, dilution and spike-recovery corrections. 

\subsection{Vessel cleaning and blank rates} 
\label{subsec:vessel_blanks}

Control of the analytical blank is key to achieving detection limits low enough to assay materials for rare-event searches. When using SpA-grade reagents for microwave digestion the dominant contribution to the blank rate is the purity of the reagent itself, yielding typically a few ppt (g/g) of $^{232}$Th and $^{238}$U in the final 20\% solution and limiting the detection limit to $>$ppb-levels in terms of sample concentration. For digestions with UpA-grade reagents desorption of impurities in the TFM digestion vessels, PP labware and the sample introduction system become important. 

We apply two levels of cleaning for plastic labware such as the PP 50 mL DigiTubes, 1 L High Density Polyethylene (HDPE) bottes and 6 mL  auto-sampler vials. Unlike the TFM digestion vessels, they do not come into contact with concentrated acids at elevated temperatures and we have found that using new labware that has simply been rinsed three times with DI water (and left to dry in the LFU if being used for weighing) is sufficient for most purposes as typical U/Th BECs seen in calibrations and samples prepared are below $< 30$ ppq. For samples requiring lower blank rates we leach all labware in 20, 10 or 5\% v/v SpA nitric acid for 24 hours prior to use, where the concentration is chosen to match the expected use. An example of this was shown in Table~\ref{tab:Ar_checks}, where we saw a halving of the blank rate for 5\% UpA blanks prepared in DigiTubes that were first leached in 10\% SpA HNO$_{3}$ for 24 hours.      

As they hold concentrated reagents at elevated temperatures contamination from the digestion vessels is of particular concern. Following sample digestion each TFM vessel and lid is rinsed a minimum of three times with DI water and then, after drying, is put through a decontamination microwave program. For routine cleaning this is a 15 minute ramp and 15 minute hold at 180$^{\circ}$C with 5~mL of HNO$_{3}$ SpA-grade and 5 mL of DI water. If a further reduction in blank rate is required the process is repeated with higher purity reagents, longer hold time or by matching the reagents to those being used for the subsequent digestion. For example, for the open vessel digestions of Ti wire in \S\ref{sec:ti_digestions} the vessel cleaning routine included an additional 2 mL of hydrofluoric acid.   

We now compare process blank rates for vessels pre-cleaned using different purity HNO$_{3}$, first with SpA-grade and then with doubly-distilled SpA. Table~\ref{tab:blank_rates} shows process blank rates measured following closed vessel digestion of 10~mL of UpA-grade nitric acid on a typical microwave routine (15 minute ramp and 20 minute hold at 210$^{\circ}$C). There is a clear improvement when switching to doubly-distilled SpA with the average blank rates coming down to $<5$~ppq for $^{232}$Th and $<10$~ppq for $^{238}$U. More importantly, the variability of the blank rate reduces significantly with standard deviations going from $>10$~ppq to $<3$~ppq. 

Taking $3\times$ the standard deviation on the blank rates in Table~\ref{tab:blank_rates} as a guide to the achievable detection limits\footnote{Actual detection limits will depend on the particular sample being assayed and factors such as instrument stability, matrix effects and total integration time per mass.} and accounting for a hypothetical 500-fold dilution (0.1~g digested sample diluted to 50 mL final solution) this implies detection limits of around 20 ppt g/g for vessels pre-cleaned with SpA HNO$_{3}$ and below 5 ppt g/g when using vessels cleaned with doubly-distilled-SpA.   

\begin{table}[h]
\centering
\caption{Comparison of process blank rates (in ppq g/g of 50 mL final diluted solution) for closed vessel microwave digestion with 10 mL HNO$_{3}$ (ROMIL UpA-grade) for 20 minutes at 210$^{\circ}$C. Blank rates were measured twice for four vessels following the same microwave cleaning procedure (20 minutes at 180$^{\circ}$C with 50\% HNO$_{3}$) but with different purity HNO$_{3}$: once with SpA-grade HNO$_{3}$ and once with doubly-distilled SpA-grade HNO$_{3}$. The large relative uncertainties for the distilled-SpA pre-cleaned blanks indicates their proximity to the instrument detection limits ($\sim$10 ppq for this batch).}
\medskip
\begin{tabular}{ r | r | r }
  \multicolumn{3}{r}{Blank rate: ppq g/g diluted solution} \\
  \multicolumn{1}{c}{} & $^{232}$Th & $^{238}$U \\
  \hline
  \multicolumn{3}{c}{Vessel pre-cleaned with SpA} \\
  V1 & $ 19 \pm 9 $ & $ 32 \pm 6$ \\
  V2 & $ 16 \pm 5$ & $8 \pm 4$ \\
  V3 & $ 11 \pm 7$ & $40 \pm 12$ \\
  V4 & $ 38 \pm 12$ & $33 \pm 7$ \\
  Avg. $\pm$ Std.Dev & $21 \pm 12$ & $29 \pm 14$ \\
  \hline  
  \multicolumn{3}{c}{Vessel pre-cleaned with distilled-SpA} \\
  V1 & $3.8 \pm 7.2$ & $8.2 \pm 5.7$ \\
  V2 & $0.4 \pm 4.3$ & $7.4 \pm 2.5$ \\
  V3 & $6.1 \pm 6.1$ & $9.0 \pm 8.4$ \\
  V4 & $3.6 \pm 3.5$ & $9.1 \pm 6.8$ \\
  Avg. $\pm$ Std.Dev & $3.5 \pm 2.3$ & $8.7 \pm 1.1$ \\
\end{tabular}
\label{tab:blank_rates}
\end{table}%

The reduced blank rates demonstrated with the double-distilled SpA pre-cleaning are at or below instrument detection limits when running the ICP-MS in a General Purpose plasma mode with a standard tune for high-mass sensitivity and stability. It is reported that with a custom tuning the Agilent 7900 can achieve $>1000$ CPS/ppt response with instrument detection limits of 1.3 ppq with BEC of 0.48 ppq for a $^{238}$U calibration~\cite{ref:agilent7900release}. In these circumstances further steps to reduce the process blank rate would be motivated, such as using only fluorinated polymers (FEP, PFA) for all labware and the use the traceCLEAN acid steam cleaning system.

\subsection{$^{230}$Th and $^{233}$U spikes}

Isotopic reference materials for $^{230}$Th and $^{233}$U were purchased from the Institute for Reference Materials and Measurements (IRMM)\footnote{Located in Geel, Belgium.} for use as spikes in isotopic dilution analysis. Each reference material was supplied in a glass ampule containing 5 mL of chemically stable nitric acid and was subsequently diluted using 5\% v/v UpA nitric acid into pre-conditioned\footnote{Leached for 48 hours in 20\% v/v SpA nitric acid.} FEP bottles to make $\sim 100$ mL stock solutions. These stock solutions are then further diluted as needed to make $\sim$ppb g/g working solutions from which the 1--100~ppt typical spike and calibration solutions are made up.

Tables~\ref{tab:JRC-IRMM_Th230_spike_prep} and \ref{tab:JRC-IRMM_U233_spike_prep} show calculated concentrations for the stock, working and a final spike solution based on certified values for the $^{230}$Th (IRMM--061~\cite{ref:IRMMTh230refsheet}) and $^{233}$U (IRMM--051~\cite{ref:IRMMU233refsheet}) spikes, respectively. To confirm the isotope ratios as well as the overall dilution scale the spike solutions were measured using an ICP-MS run with external calibrations (matrix matched) and show good agreement for all isotopes. 

\begin{table}[h]
\centering
\caption{IRMM--061 $^{230}$Th spike isotope concentrations based on certified values~\cite{ref:IRMMTh230refsheet} for the various stages of dilution. ICP-MS measured concentrations of the final spike solution using an external $^{232}$Th calibration are in good agreement.}
\medskip
\resizebox{0.7\linewidth}{!}{
\begin{tabular}{ r | c | c }
  \multicolumn{1}{c}{} & $^{230}$Th & $^{232}$Th \\
  \hline
  Ampule [ppm] & 0.5691 & 0.00086  \\
  Stock [ppb] & 30.77 & 0.047 \\
  Working [ppt] & 214.2 & 0.324 \\
  Spike [ppt] & 38.49 & 0.058 \\
  \hline
  Meas. [ppt] & $ 39.2 \pm 0.3$ & $0.06 \pm 0.01$ \\
\end{tabular}
}
\label{tab:JRC-IRMM_Th230_spike_prep}
\end{table}%

\begin{table}[h]
\centering
\caption{IRMM--051 $^{233}$U spike isotope concentrations based on certified values~\cite{ref:IRMMU233refsheet} for the various stages of dilution. ICP-MS measured concentrations of the final spike solution using an external $^{238}$U calibration are in good agreement.}
\medskip
\resizebox{\linewidth}{!}{
\begin{tabular}{ r | c | c | c | c | c }
  \multicolumn{1}{c}{} & $^{233}$U & $^{234}$U & $^{235}$U & $^{236}$U & $^{238}$U \\
  \hline
  Ampule [ppm] & 2.354 & 0.0221 & 0.0052 & 0.0006 & 0.0197 \\
  Stock [ppb] & 142.9 & 1.339 & 0.315 & 0.036 & 1.197 \\
  Working [ppt] & 1436 & 13.46 & 3.16 & 0.36 & 12.03 \\
  Spike [ppt] & 46.8 & 0.439 & 0.103 & 0.012 & 0.392 \\
  \hline
  Meas. [ppt] & $ 48 \pm 3 $ & $ 0.44 \pm 0.04$ & $0.1 \pm 0.02$ & $< 0.1$ & $0.38 \pm 0.02$ \\
\end{tabular}
}
\label{tab:JRC-IRMM_U233_spike_prep}
\end{table}%

The low spike ratios of $n(^{232}\mathrm{Th})/n(^{230}\mathrm{Th}) < 0.2$\% and $n(^{238}\mathrm{U})/n(^{233}\mathrm{U}) < 1$\% and the proximity of typical low-background materials to instrument detection limits means that for many materials full isotope dilution analysis to determine $^{232}$Th and $^{238}$U reduces to a simple spike recovery correction. 

\subsection{$^{232}$Th and $^{238}$U recovery for IAEA--385 Irish Sea Sediment CRM}

To verify $^{232}$Th and $^{238}$U recovery in a realistic sample matrix four samples of reference material IAEA--385~\cite{ref:IAEArefsheet} were processed for comparison to certified values. IAEA--385 is a powder made from sediment collected from the Irish Sea in 1995 which was ground, sieved and homogenized and then certified for a range of radionuclides. Consisting mainly of Si (160 mg/g), Ca (55 mg/g), Al (45 mg/g), Fe (31 mg/g) and K (18 mg/g), it represents a complex matrix requiring multi-acid digestion. The certified values of 33.7 Bq/kg $^{232}$Th and 29.0 Bq/kg $^{238}$U translate to concentrations of 8.3 ppm and 2.3 ppm respectively. 

Samples were digested using the ETHOS--UP microwave following a method developed for soil samples with Analytix. Up to 0.2 g of sample was weighed into each TFM vessel and then 8 mL HNO$_{3}$ (ROMIL SpA) + 2 mL HF (Fisher TraceMetal) + 2 mL H2O2 (Merck Suprapur) was added. The vessels were left for 10 minutes before running through a modified ETHOS--UP Sea Sediment method (SK--Environmental--026): 15 minute ramp to 200$^{\circ}$C, then hold for 20 minutes, then fan cool for 40 minutes. Following digestion the vessel contents were decanted into 50 mL DigiTubes and diluted up to a total volume of 50 mL using deionized water, a dilution of around 250. The digested solution was a clear light green/blue liquid. Due to the high $>$ppm levels of contamination expected a further factor 100 dilution was performed to bring final solution concentrations down to $O$(100 ppt) to avoid potential contamination of the sample introduction system and ICP-MS itself. In addition to the four IAEA-385 samples three process blanks were prepared. Samples were not spiked with $^{230}$Th and $^{233}$U as this would have required a significant quantity of each stock solution, with an associated potential for contamination, to be added to each vessel to yield comparable concentrations to those expected for $^{232}$Th and $^{238}$U.  

Twice diluted samples and process blanks were processed in a single batch with the Agilent 7900 running in General Purpose plasma mode and auto-tuned for high-mass sensitivity. 20, 75, 150 and 300 ppt external calibrations were used yielding instrument sensitivities of 350 CPS/ppt $^{232}$Th and 351 CPS/ppt $^{238}$U with calibration blank rates $< 1$ ppt. Measured concentrations in ppm g/g IAEA powder are shown in Table~\ref{tab:IAEA_recovery}. These are corrected for sample dilution and include a sub-dominant process blank rate subtraction ($< 0.01$ ppm). Individual results for each sample are shown as well as their average and standard deviation. The equivalent Bq/kg specific activity can be compared to the IAEA certified values (95\% confidence intervals in square brackets). For $^{238}$U the measured specific activity is within 2\% of the certified value whereas for $^{232}$Th there is a $\sim$20\% deficit indicating a partial recovery. 

\begin{table}[h]
\centering
\caption{ICP-MS analysis of certified reference material IAEA-385, sediment from the Irish Sea. Measured ppm g/g values are corrected for sample dilution and include a sub-dominant process blank subtraction ($< 0.01$ ppm). For the final values we take the average and standard deviation of all four samples, convert to Bq/kg and compare with certified values and their 95\% confidence intervals~\cite{ref:IAEArefsheet}.}
\medskip
\resizebox{\linewidth}{!}{
\begin{tabular}{ r | b{1.35cm} | b{1.35cm} | c c }
  \multicolumn{1}{c}{} & \multicolumn{1}{c}{} & \multicolumn{1}{c|}{} & \multicolumn{2}{c}{ppm g/g Sea sediment} \\
   & Sample mass (g) & Sample conc. \% & $^{232}$Th & $^{238}$U \\
  \hline
   \#1 & 0.092 & 0.0018 & $ 7.27 \pm 0.14 $ & $ 2.41 \pm 0.02 $ \\
   \#2 & 0.180 & 0.0034 & $ 6.31 \pm 0.08 $ & $ 2.33 \pm 0.02 $ \\
   \#3 & 0.176 & 0.0035 & $ 5.95 \pm 0.10 $ & $ 2.37 \pm 0.02 $ \\
   \#4 & 0.188 & 0.0029 & $ 6.35 \pm 0.06 $ & $ 2.41 \pm 0.04 $ \\
   Avg. & --- & --- & $ 6.47 \pm 0.56 $ & $ 2.38 \pm 0.04 $ \\
\hline
   \multicolumn{3}{r}{Equiv. Bq/kg:} & $ 26.3 \pm 2.3 $ & $ 29.4 \pm 0.5 $ \\
   \multicolumn{3}{r}{c.f. IAEA certified:} & $33.7$ [32.8--33.9] & $29$ [28--30] \\
\end{tabular}
}
\label{tab:IAEA_recovery}
\end{table}%

The partial recovery of $^{232}$Th could be due to a number of factors such as the formation of insoluble flouride species formed by the reaction of Th and hydrofluoric acid or the over-dilution of the digested samples such that $^{232}$Th is no longer stable in the solution. Further measurements with a boric acid (H3BO3) complexation step and with judiciously chosen $^{230}$Th and $^{233}$U spikes would likely resolve the source of partial recovery. For now an additional 20\% systematic uncertainty is assumed for $^{232}$Th  measurements. 

\section{Results for typical low-background materials} 
\label{sec:ti_and_inconel_assays}

To illustrate both a high-sensitivity assay using open digestion and a more routine material requiring closed-vessel digestion in a multi-acid mixture we present results for two low-background materials that were assayed as part of the LZ materials screening campaign: an open vessel digestion of small drilling burrs taken from a slab of ultra-low background Ti and a closed-vessel digestion of a nickel-chromium-alloy wire. 

\subsection{Open digestion: ultra-pure Ti chips}
\label{sec:ti_digestions}

TIMET HN3469 is a single 15,000 kg slab of ultra-radiopure ASTM Grade 1 titanium procured to make the LZ cryostat following an extensive radio-assay campaign~\cite{Akerib:2017iwt}. A sample of $\sim 2$ g of burrs drilled from a test coupon, taken from the middle of TIMET HN3469, were provided for rapid ICP-MS analysis for the purposes of quality control. For this coupon the inferred values of  $^{238}$U and $^{232}$Th as determined through gamma-spectroscopy measurements with high purity germanium (HPGe) detectors were $^{238}$U = 2.8 $\pm$ 0.15~mBq/kg and $^{232}$Th $<$ 0.2~mBq/kg. The small sample mass, low activity and fast turnaround makes this a canonical use-case for ICP-MS.  

Prior to digestion samples were sonicated twice in high-purity IPA (Fisher Chemical Optima grade $> 99.9\%$), rinsed in DI water and then etched in a 2\% v/v HF and 2\% v/v HNO$_{3}$ solution. The etching process consumed around 30\% of the mass, after which samples were rinsed in DI water and left to dry overnight in the ISO Class 5 LFU. At each stage a small amount of the sample was set aside to gauge the effectiveness of each step. Four samples with masses between 0.1 and 0.2 g were prepared for digestion: samples \#1 and \#2 had been both cleaned and etched, sample \#3 was only cleaned and sample \#4 was neither cleaned nor etched.

Samples were weighed into open TFM vessels (pre-cleaned in the ETHOS-UP with SpA-grade HNO$_{3}$) and spiked with $\sim 1.5$ mL of 50 ppt $^{230}$Th and $^{233}$U solution, to give $\sim0.75$ ppt in diluted samples. In addition to the sample vessels three process blanks were prepared and spiked. To digest, 10 mL of a $\sim$10\% v/v HF and $\sim$10\% v/v HNO$_{3}$ mixture was added to each vessel. The vessels were left for an hour (visually, samples were dissolved within 5 minutes) prior to a two-stage dilution with DI water to bring the total sample concentration to below 0.2\%, the maximum recommended TDS for the ICP-MS. For both etching and digestion only ROMIL UpA-grade HF and doubly-distilled SpA HNO$_{3}$ were used.

All diluted samples and blanks were processed in a single ICP-MS batch in General Purpose plasma mode and auto-tuned to high-mass sensitivity. External calibrations of 0.05, 0.25, 1.0 and 5.0 ppt established instrument sensitivities of 72 CPS/ppt $^{230}$Th, 71 CPS/ppt $^{232}$Th, 77 CPS/ppt $^{233}$U and 77 CPS/ppt $^{238}$U. The low CPS/ppt responses were due to buildup on the sampling cone following earlier runs with high TDS. 

Results are shown in Table~\ref{tab:Ti_assay} where the Ti sample ordering corresponds to their order in the batch run. Process blank rates were below the instrument detection limits and show full spike recovery for both $^{230}$Th and $^{232}$Th. Spike recovery drops significantly throughout the run, consistent with the response from the $^{193}$Ir and $^{209}$Bi internal standards. Results for ppt g/g of Ti are corrected for spike recovery and sample mass/dilution on a sample-by-sample basis. All measurements were consistent with results from HPGe assays with the exception of $^{232}$Th in sample \#4, which showed an elevated activity consistent with the introduction of surface contamination during processing and handling. The uncertainties on these measurements were not process blank rate limited and it is expected that with a more typical instrument sensitivity ($\sim 400$ CPS/ppt, achievable after cleaning) measurements down to 10~ppt g/g of can be achieved.

\begin{table*}[!htbp]
\centering
\caption{ICP-MS analysis of TIMET HN3469--M Ti chips provided by the LZ collaboration. Samples were digested in open TFM vessels using 8 mL DI water + 1 mL HF + 1 mL HNO$_{3}$ (both ROMIL UpA-grade). $^{232}$Th and $^{238}$U ppt values are per gram of sample mass and include process blank and spike recovery corrections. Ti samples \#1 and \#2 were cleaned and etched prior to digestion, sample \#3 was cleaned but not etched and sample \#4 was neither cleaned or etched. External calibrations established the instrument response at the start of the run to be 71 CPS/ppt $^{232}$Th and 77 CPS/ppt $^{238}$U. Throughout the ICP-MS run this response degraded due to buildup on sampling and interface cones. Per sample spike recovery was used to correct the external calibrations for this time dependence plus any matrix or digestion efficiency effects.}
\medskip
\resizebox{\textwidth}{!}{
\begin{tabular}{ r | b{1.3cm} | b{1.3cm} | r r | r r | r r | r r }
  \multicolumn{1}{c}{} & \multicolumn{1}{c}{} & \multicolumn{1}{c|}{} & \multicolumn{2}{c|}{ppq g/g solution} & \multicolumn{2}{c|}{spike recovery \%} & \multicolumn{2}{c|}{ppt g/g of Ti} & \multicolumn{2}{c}{mBq/kg of Ti} \\
  & Sample mass (g) & Sample conc. \% & $^{232}$Th & $^{238}$U & $^{230}$Th & $^{233}$U & $^{232}$Th & $^{238}$U & $^{232}$Th & $^{238}$U \\
  \hline
Proc. Blank & --- & --- & $6 \pm 10$ & $10 \pm 11$ & $97 \pm 12$ & $98 \pm 10$ & --- & --- & --- & --- \\
  Ti \#2 & 0.202 & 0.157 & $ 56 \pm 24 $ & $227 \pm 49$ & $90 \pm 8$ & $85 \pm 16$ & $36 \pm 18$ & $163 \pm 35$ & $0.15 \pm 0.08$ &  $2.0 \pm 0.4$\\
Ti \#1 & 0.129 & 0.161 & $ 37 \pm 21 $ & $ 211 \pm 37 $ & $ 78 \pm 6 $ & $70 \pm 7$ & $25 \pm 19$ & $180 \pm 30$ & $0.10 \pm 0.08$ &  $2.2 \pm 0.4$\\
  Ti \#3 & 0.215 & 0.191 & $30 \pm 30$ & $140 \pm 30$ & $43 \pm 7$ & $39 \pm 4$ & $32 \pm 43$ & $175 \pm 42$ & $0.13 \pm 0.17$ &  $2.2 \pm 0.5$\\
Ti \#4 & 0.091 & 0.180 & $94 \pm 27$ & $139 \pm 35$ & $41 \pm 6$ & $36 \pm 7$ & $119 \pm 38$ & $197 \pm 49$ & $0.48 \pm 0.16$ &  $2.4 \pm 0.6$\\
\end{tabular}
}
\label{tab:Ti_assay}
\end{table*}%

\subsection{Closed vessel microwave digestion: Ni-based alloy}

A 10~g piece of nickel-chromium-alloy (Inconel) wire sample was made available for assay. Before digestion the wire was cleaned using IPA-cleanroom wipes, rinsed with DI water and left to dry overnight in the LFU. Three small $\sim$0.1 g sections of Inconel were cut\footnote{Using a Knipex Chrome Vanadium steel wire cutter pre-cleaned with IPA-wipes and DI water.} and weighed directly into the digestion vessels and two process blanks were prepared. One sample (\#1) and one process blank were spiked with 1.5 mL of 50 ppt $^{230}$Th and $^{233}$U solution, to give $\sim1.5$ ppt in the diluted solution.  

Samples were digested in 6 mL HCl + 3 mL HNO$_{3}$ + 1 mL HF (all ROMIL UpA-grade) using closed vessel microwave digestion following the ETHOS-UP SK--Metal--018 Inconel routine: 20 minute ramp to 220$^{\circ}$C and then 15 minute hold at 220$^{\circ}$C. Following digestion, samples were decanted into 50 mL PP DigiTubes and diluted down to $<0.2$\% concentration by topping up to 50 mL with DI water. 

Samples were processed in a single ICP-MS batch with the instrument in General Purpose plasma mode and auto-tuned for high-mass sensitivity. External calibrations of 0.2, 1.0 and 20.0 ppt established instrument sensitivities of 238 CPS/ppt $^{230}$Th, 238 CPS/ppt $^{232}$Th, 272 CPS/ppt $^{233}$U and 272 CPS/ppt $^{238}$U. Results are shown in Table~\ref{tab:NiAlloy_assay}. Measured ppb g/g of Ni-Cr-alloy concentrations include a sub-dominant average process blank subtraction and are corrected for the spike recovery in sample \#1. Errors on individual sample measurements include the instrument statistical error, uncertainty in the process blank and the error associated with weighing the $\sim1$ g samples. The use of a single spike correction results in an additional 10\% systematic error giving final specific activities for $^{232}$Th of $4.0 \pm 0.1 (\mathrm{meas.}) \pm 0.4 (\mathrm{syst.})$~mBq/kg, and $110.9 \pm 2.1 (\mathrm{meas.}) \pm 11.1 (\mathrm{syst.})$~mBq/kg for $^{238}$U. This measurement demonstrates the  precision that can be achieved for samples with relatively high activities. The additional systematic would be eliminated by the use of appropriate quantities of spikes in all samples.

\begin{table*}[!htbp]
\centering
\caption{ICP-MS analysis of nickel-chromium-alloy (Inconel) wire sample provided by the LZ collaboration. Samples were microwave digested in 6 mL HCl + 3 mL HNO$_{3}$ + 1 mL HF (all ROMIL UpA-grade) following the ETHOS-UP SK-Metal-018 Inconel routine: 20 minute ramp to 220$^{\circ}$C and then 15 minute hold at 220$^{\circ}$C. $^{232}$Th and $^{238}$U ppb values are per gram of sample mass and include process blank and spike recovery corrections (using sample \#1). There is an additional 10\% systematic associated with the spike-correction. Results are consistent across the three samples assayed.}
\medskip
\resizebox{\textwidth}{!}{
\begin{tabular}{ r | b{1.3cm} | b{1.3cm} | c c | c c | c c | c c }
\multicolumn{1}{c}{} & \multicolumn{1}{c}{} & \multicolumn{1}{c|}{} & \multicolumn{2}{c|}{ppt g/g solution} & \multicolumn{2}{c|}{spike recovery \%} & \multicolumn{2}{c|}{ppb g/g of Ni-Cr} & \multicolumn{2}{c}{mBq/kg of Ni-Cr} \\
  & Sample mass (g) & Sample conc. \% & $^{232}$Th & $^{238}$U & $^{230}$Th & $^{233}$U & $^{232}$Th & $^{238}$U & $^{232}$Th & $^{238}$U \\
  \hline
  Proc. Blanks & --- & --- & $ 0.01 \pm 0.01 $ & $ 0.03 \pm 0.01 $ & $ 92 \pm 3 $ & $ 85 \pm 3 $ & --- & --- & --- & --- \\
  Ni-Cr \#1 & 0.096 & 0.182 & $ 1.89 \pm 0.04 $ & $ 47.1 \pm 0.6 $ & $ 106 \pm 7 $ & $ 94 \pm 6 $ & $0.97 \pm 0.02 $ & $ 27.6 \pm 0.3 $ & $ 4.0 \pm 0.1 $ &  $ 112.3 \pm 1.2 $ \\
  Ni-Cr \#2 & 0.105 & 0.188 & $ 1.97 \pm 0.11 $ & $ 48.6 \pm 0.4 $ & --- & --- & $ 0.98 \pm 0.06 $ & $ 27.6 \pm 0.02 $ & $ 4.0 \pm 0.2 $ &  $ 112.0 \pm 0.8 $\\
  Ni-Cr \#3 & 0.096 & 0.165 & $ 1.81 \pm 0.07 $ & $ 41.4 \pm 0.6 $ & --- & --- & $ 1.02 \pm 0.04 $ & $ 26.7 \pm 0.3 $ & $ 4.2 \pm 0.2 $ &  $ 108.6 \pm 1.3 $\\
\multicolumn{7}{r|}{Avg. $\pm$ Std.Dev:} & $0.99 \pm 0.03$ & $27.3 \pm 0.5$ & $4.0 \pm 0.1$ & $110.9 \pm 2.1$ \\
\end{tabular}
}
\label{tab:NiAlloy_assay}
\end{table*}

\section{Summary and Outlook}
\label{sec:summary}

The UCL ICP-MS rare-event search laboratory serves as a dedicated materials screening facility for the LZ dark matter experiment. The primary instrument is an Agilent 7900 ICP-MS with an instrument sensitivity down to $\sim10$ ppq U/Th and fitted with an inert sample introduction kit to allow direct sampling of solutions with up to $20\%$ acid concentrations, including HF. The ICP-MS is operated inside an ISO Class 6 cleanroom alongside all necessary infrastructure for sample preparation.

Straightforward cleaning, sample preparation and analysis procedures have been developed to allow high-throughput and fast-turnaround of materials with U/Th concentrations down to 10 ppt g/g of sample. These are based on microwave assisted sample digestion in pre-cleaned TFM vessels and then direct ICP-MS measurement of the diluted sample with no further chemical treatment or evaporation steps. We use $^{230}$Th and $^{233}$U spikes to correct for signal loss from a range of sources, and demonstrate the importance of their use through digestion and assay of a complex sea sediment CRM. These procedures allow a complete assay, including digestion, ICP-MS measurement and analysis, to be completed in a single day. 

We have evaluated the impact of reagent purity on U/Th background levels and found that for most routine assays it is possible to use doubly-distilled ROMIL-SpA nitric acid as an almost equivalent and cheap alternative to ultra-pure ROMIL-UpA grade acid. In addition, it was shown that switching ICP-MS Ar carrier gas purity from research-grade (N5.5) to zero-grade (N5.0)  causes no significant increase in U/Th background level given current instrument sensitivity and labware cleaning protocols. 

Results were presented for two typical low-background materials assayed as part of the LZ materials screening campaign: an open digestion of ultra-pure Ti demonstrating a $\sim30$ ppt g/g $^{232}$Th measurement and showing consistency with HPGe results for both  Th and U; and a closed vessel digestion of a nickel-chromium-alloy wire requiring a multi-acid mixture as an example of a high-precision measurement of a ppb-level sample. These demonstrate the suitability of the procedures to screen the majority of materials for the current generation low-background experiments. Over 100 assays have been performed for LZ with this facility already as part of the experiment's construction phase. Results from these measurements will be presented by the collaboration in the future.

Looking ahead, future low-background experiments, such as so-called `Generation-3' dark matter, and next generation $0\nu\beta\beta$ searches will require backgrounds and consequent radio-assay capability at least an order of magnitude better than present searches. ICP-MS, with both requisite sensitivity to direct $^{232}$Th and $^{238}$U measurements and high throughput, will be a key technique, complementing gamma-spectroscopy, radon emanation, and material surface activity measurements in the radio-assay campaigns of all such future experiments. Moreover, sensitivity can be enhanced further. Improved cleanliness procedures and matrix separation techniques, see for example~\cite{LaFerriere:2014rva}, can be applied to concentrate U/Th and boost $^{232}$Th and $^{238}$U signal to sub-ppt sensitivity. 

\section*{Acknowledgements}
This work was supported by the U.K. Science \& Technology Facilities Council (STFC) under award numbers ST/M003981/1, ST/L006170/1, and ST/M006891/1, as well as an STFC Impact Acceleration Award (University College London No. 156822). We appreciate support from the U.K. Royal Society for travel funds under the International Exchange Scheme (Award No. IE141517). We are grateful for departmental support for laboratory and clean room infrastructure for this facility. The authors would also like to thank Jerry Busenitz and Yue Meng (University of Alabama), Kevin Lesko (Lawrence Berkeley National Laboratory), Doug Leonard (Centre for Underground Physics, S. Korea), and Eric Hoppe and Isac Arnquist (Pacific Northwestern National Laboratory) for useful discussions and assistance with ICP-MS techniques and protocols. We thank L. Reichhart for her work at UCL during commissioning of the ICP-MS instrument. We thank Duncan Rowe and Jack Cartwright from Analytix Ltd. for their collaboration in sample preparation methods, and Sonia North and Raimund Wahlen from Agilent Technologies for their support with instrument operations. Finally, we wish to thank all of our collaborators in the LZ dark matter Experiment for their ongoing support.   

 
\bibliography{mybibfile}

\end{document}